# The Patient/Industry Trade-off in Medical Artificial Intelligence

**Rina Khan[1], Annabelle Sauve[2], Imaan Bayoumi[3], Amber L. Simpson[1,4], Catherine Stinson[1,5]**

[1] School of Computing, Queen's University  [2] LGI Healthcare Solutions  [3] Department of Family Medicine, Queen's University  [4] Department of Biomolecular Sciences, Queen's University
[5] Department of Philosophy, Queen's University

**Abstract**

Artificial intelligence (AI) in healthcare has led to many promising developments; however, increasingly, AI research is funded by the private sector leading to potential trade-offs between benefits to patients and benefits to industry. Health AI practitioners should prioritize successful adaptation into clinical practice in order to provide meaningful benefits to patients, but translation usually requires collaboration with industry. We discuss three features of AI studies that hamper the integration of AI into clinical practice from the perspective of researchers and clinicians. These include lack of clinically relevant metrics, lack of clinical trials and longitudinal studies to validate results, and lack of patient and physician involvement in the development process. For partnerships between industry and health research to be sustainable, a balance must be established between patient and industry benefit. We propose three approaches for addressing this gap: improved transparency and explainability of AI models, fostering relationships with industry partners that have a reputation for centering patient benefit in their practices, and prioritization of overall healthcare benefits. With these priorities, we can sooner realize meaningful AI technologies used by clinicians where mutually beneficial impacts for patients, healthcare providers, and industry can be realized.

## 1 Background

Partnerships between artificial intelligence (AI) and medical communities have led to promising advances. In computer-aided detection, a subfield of AI, impressively high accuracy is being achieved in the automatic detections and diagnosis many cancers including colorectal [1], breast [2], lung [3], and brain [4]. Many attempts at COVID-19 detection in CT and X-ray scans have been made with some success [5]. Interpretable multi-objective models have been shown to jointly predict mortality and hospitalization, proving how transparent machine learning tools can offer clinically significant benefits [6]. Language-based approaches for structuring electronic medical records are increasingly prominent [7]. These types of applications aim to extract meaningful information from data and present it in a format that is easy to process and apply.

The increasing availability of data in combination with powerful AI algorithms and widespread availability of advanced computational hardware offers many benefits to both healthcare systems and patients. The main aims of these collaborations are that health industries achieve "increased productivity with decreased costs, as well as reductions in medical error" [8]. As a result, patients may benefit from improved individualized care with greater efficiency and lower cost.

As encouraging as this seems, there are many societal and ethical barriers to consider when integrating AI applications into healthcare. These include conforming with legislations such as General Data Protection Regulation (GDPR), protecting privacy of patients and ensuring ethical and fair use [9]. Another widely held concern is the risk of biases in AI applications, which include racial, gender, and age biases, among others [10].

This paper examines some of the challenges of producing clinically meaningful work, considering the perspectives of patients, practitioners, and funding industries, and offers direction for how to improve the prospects of translating medical AI into clinical practice. We look at the patient and industry trade-off in greater detail. We analyze factors that contribute towards patient benefit, looking at clinically relevant metrics in AI development, randomized control trials in relation to AI, validation of results with real-word evidence and involving patients and physicians in the development process. We then consider how industry benefit can be achieved while keeping patient benefits in mind. This includes building explainable systems grounded in trust, prioritizing long-term reputation building with health-care providers, contributing towards healthcare benefits, and being transparent about funding.

## 2  The patient/industry trade-off

The primary goal of medical research should be to improve human health, however, additional motives can distort research incentives. A 2009 article by Chalmers and Glasziou reported that over 85% of health research, and their associated funds, go to waste [11]. This waste has four main causes: investigating irrelevant research questions, using inappropriate designs and methods, over-reporting positive results while under-reporting negative results, and producing untranslatable research. This raises the question of whether the goal of improving health is adequately considered when researchers draft proposals.

The blame cannot solely be put on researchers, however, because research funders play a major role in deciding what research is conducted, how, and by whom. Private industry sponsors see research as an investment and "rarely ask for a systemic assessment of the need for the proposed research" [12]. Private industry funding will have to be clinically beneficial due to regulations around healthcare research, but such intentions are undermined by a focus on short term goals and "maximum profitable return on investment" [12]. Academic institutions engaging with such funders are forced to operate as profit generating businesses [12]. Profit motive and shorter time frames to complete industry funded studies which leave out long term patient impact negatively affect the actual benefit in relation to patient needs. If most medical research is motivated by profit and ends up going nowhere in terms of clinical translation, patients' needs and the public good are being neglected.

In 2018, P. A. Keane and E. J. Topol coined the term 'AI Chasm', to name the disconnect between the accuracy of algorithms used in medical AI, and their real-world clinical significance [13]. The authors noted that algorithms usually do not generalize well outside of the niche population on which they were trained and would need remodeling to be efficient in clinical settings. Despite many promises made about how AI might be applied to medicine, many published studies do not grapple with how to cross the AI Chasm and never get translated into the clinical space [14]. A balance of beneficial patient outcomes, ease of clinical implementation and provider support are necessary for successful use of AI medical diagnostic tools [15]. Successful adaptation of medical AI into clinical practice needs to be given higher priority.

*Table 1* Study requirements to ensure maximization of patient benefit during clinical integration of AI.

| Requirements | Relevant ethical principles | Issue | Example |
|---|---|---|---|
| Clinically relevant metrics | *Beneficence:* Clinically relevant metrics prioritize patient benefit.<br>*Trust:* Clinically relevant metrics provide a higher degree of trust to the healthcare provider that will be using the system | Metrics currently used in AI research do not represent impact on a patient. | A model that proposes treatment options includes a metric for patient benefits and status. |
| Randomized clinical trials | *Beneficence:* Clinical trials prove efficacy and effectiveness with respect to patient outcome.<br>*Non-maleficence:* Clinical trials are meant to also to observe toxicity. This prevents future harm from entering real-world settings.<br>*Trust:* Patients and physicians are more likely to trust a system if its efficacy and effectiveness have been established in a controlled setting. | Clinical trials are necessary to determine the efficacy of a drug treatment within a controlled setting. | A diagnostic AI model is tested in health clinics where a physician can observe the performance and assess the real-world accuracy. |
| Validation with real world evidence | *Beneficence:* Longitudinal studies are used to validate a system's results. This promotes patient benefit.<br>*Justice:* Longitudinal studies validate systems on a wide variety of participants, promoting equality.<br>*Trust:* Patients and physicians are more likely to trust a system if it has been validated with external data. | Baseline comparisons made with real-world data collected in a natural setting are needed to validate the results of a system. | A prospective study is conducted on patients at risk of developing a disease. An AI system is developed in parallel and makes predictions about the participants. The predictions are compared to the ground truth data from the prospective study. |
| Patient and physician involvement | *Autonomy:* Involving patients and physicians in the process promotes awareness, allowing for informed decision-making, and self-governance.<br>*Trust:* Patients and physicians are more likely to trust a system they were involved in developing. | Patients and physicians are directly participating in AI studies by sharing their data and collaborating. | Patients collaborating with researchers can determine long term outcomes most beneficial to them. This can lead to development of patient important outcomes that can then be prioritized. |

Here, we examine the trade-off in more detail.

## 2.1 Patient Benefit

The patient is the most important stakeholder in medical research. They receive the resulting treatment and should always benefit from the outcome of medical research. While most medical AI studies are ostensibly driven by the goal of offering better healthcare, they often lack key requirements for clinical translation including clinically relevant metrics, clinical trials, baseline comparisons, and patient and physician involvement. This hampers the translation of medical AI into clinical settings, ultimately hindering the main goal of improving patient care. Table 1 summarizes these key requirements, relevant ethical principles from the Belmont Report [16] and examples.

### 2.1.1 Clinically relevant metrics and randomized control trials

In medicine, there are many ways of calculating clinical outcomes and patient benefits. There is a body of research that argues for prioritizing outcomes that are relevant to patients and having patients more involved in assessing their own risks and benefits [17]. Physicians bring their own biases to making such decisions, such as being more inclined towards clinical benefits at the risk of psychological harm for the patient [17]. Patient important outcomes (PIOs) are important metrics in quantifying how patients weigh benefits against risks. For example, a patient may value their quality of life and ability to perform tasks over facing other health risks. While patient benefit drives all clinical metrics, there is a clear distinction between clinician deciding on what is beneficial to the patient versus shared decision making with both the clinician and the patient. While there are benefits and risks to both approaches, clinical health care is moving towards centering patients and providing them with greater agency in their care [17],

However, in clinical trials, surrogate metrics that are quick and easy to collect are frequently selected over PIOs [18]. Randomized control trials (RCT) related to diabetes only used PIOs as a primary outcome in 18% of trials [18]. PIOs are not commonly used as relevant metrics when assessing medical AI models in RCTs. For example, AI models are easier to train on cholesterol levels as a surrogate for diabetes risk or the downstream effects on patients' lives [18], as this is easy to measure and clinicians can generate large amounts of data for AI model training. PIOs such as quality of life, risk of major morbid events (strokes, amputation, vision loss, delayed wound healing, pain and loss of ability to perform functional tasks) [18] are far more difficult to measure and require longer clinical trials, which make them difficult to fit into the typical lifecycle of AI development. Metrics used in the assessment of algorithms in RCTs have been found to be poor representations of patient benefit [14].

In AI research, the assessment of algorithms typically depends on mathematical concepts. Common techniques are to calculate the success of an algorithm in terms of model accuracy, the area under the receiver operating characteristic curve (AUROC), or mean squared error (MSE) [19]. While these measures may be meaningful for computer scientists, they do not directly address whether the outcome of the model is significant and beneficial for the patient [20]. Instead, they only address the certainty of a given prediction or result. For example, AUROC is used in classification problems to analyze the performance of a classifier. It is high (nearing a score of 1) when the model correctly classifies, and low when it does not. The main disadvantage of this metric is that it does not represent gains and losses for individual patients, instead evaluating performance over all patients at once [21]. More limiting is that these metrics require ground truth to be known for an individual patient, which is of course not available during routine care. Accuracy and MSE measures are similar in that they review the average or mean performance instead of individual impact.

Therefore, additional metrics that concern individual patients should be examined during the development phase. For example, a model that proposes treatment options should include computational metrics that provide information on the effectiveness of a treatment for a given patient but also include a measure for impact on a patient's survival that examines health status and life expectancy [22]. In cancer trials, the use of overall survival rather than the less relevant progress-free survival is thought to address this gap, though this solution isn't perfect as many more patients will be needed in trials [23]. Additionally, clinicians unfamiliar with the current computational metrics and the methods used to generate the results may not be able to correctly interpret the outcome if it is not accompanied by relevant patient-specific clinical metrics. Without clear evidence and explanation of an algorithm's output, the model cannot be properly employed by a physician or contribute to a patient's well-being.

There is often a lack of robustness in the testing of AI models in relation to clinical use. RCTs are necessary to prove the efficacy of a model. There are diagnostic tools that are intended to augment clinical decision-making, rather than as autonomous or semi-autonomous decision-making tools, and such tools can be continuously learning [24]. Randomized controlled trials for such continuous learning systems are currently lacking, and there are pressing challenges in their implementation, such as assessing optimization, redundancy, accuracy and optimal clinical environment [24]. For example, one of the earlier medical diagnostic decision supports (MDDS) systems, INTERNIST-1, was a rule-based expert system developed by coding the knowledge of a physician into a machine using a logic-based programming language [25]. This system would evaluate a patient's information, output possible diagnoses, and ask further questions to the user if the results were inconclusive. Since INTERNIST-1 was developed to provide a tool for physicians to help them "be more precise in characterizing the patient's illness", a randomized trial, in this case, was not necessary, because the tool was essentially replacing the literature search that a physician would have performed to uncover a diagnosis [25]. In contrast with MDDS systems, which exist to provide additional information to clinicians, some medical AI systems are able to support diagnostic decisions. For example, the Food and Drug Administration defines a computer-aided triage system as a tool that reduces a clinician's workload by reviewing a radiology image and identifying whether the patient requires further professional review or not [26, 27]. Randomized clinical trials for this type of AI are essential as an error made during the triage phase could prevent patients from obtaining the care they need.

### 2.1.2 Validation with real world evidence

Real world evidence (RWE) is clinical evidence generated from sources of real-world data (RWD) derived from routine care including electronic health records, wearable devices, etc. RWE provides complementary information to clinical trials, and inclusion and exclusion criteria test models within very specific population. Pragmatic clinical trials have been implemented to measure clinical success over time by conducting trials in real settings. In a traditional prospective study, researchers follow a group of participants over a period of time and observe how their unique characteristics affect a defined outcome (i.e. appearance or development of a disease).

A similar concept could be applied to AI prediction studies (for example, diagnosis or prognosis prediction) by comparing the natural observation outcomes with what the system had predicted. This would allow the researchers to see how effective their algorithm is in comparison to real-life patient trajectories. A retrospective study is similar except that information about an individual's past is analyzed instead of future information being collected. These studies require extensive continuous data which most healthcare facilities already collect through the form of electronic medical records. For AI research, it would be useful to simulate patients and outcomes and confirm the effectiveness of a system by comparing it to decisions that were previously made in a clinic by a physician. Prospective and retrospective studies are done in AI healthcare research and have shown to be effective [28]. The recurring theme in both types of studies is the final comparison. Conclusions drawn from an algorithm should be regularly validated in clinical settings to obtain their real-life accuracy.

A major reason why this sort of validation of predictive algorithms is necessary is that the EMRs and other medical records that comprise the training data are designed with clinical care in mind, not research, so the data can be very messy [29]. Clinicians may describe things very briefly or use acronyms which are not standardized. The records may be incomplete, especially if a patient has multiple concerns. The physician may forget to document something or may deem it to not be very consequential. As long as these types of longitudinal studies are not part of the typical workflow of machine learning applications to healthcare, claims that an AI model has higher accuracy than real physicians will not be credible, and the associated models cannot be assumed to be robust enough to undergo clinical translation.

### 2.1.3 Patient and physician involvement

Public involvement is a practice in healthcare that has many proponents and has been increasing over the recent years [30]. Proponents argue that public involvement allows patients to have input in various aspects of public health, such as how services are delivered and how research is carried out [31]. Since the patient is an important stakeholder in healthcare, they must be involved in the development of health AI. This would better align development of healthcare AI to the goals of patient-centered care, which centers patient's priorities, using tools such as patient-important outcomes as metrics. Giving patients the chance to incorporate their lived experience in healthcare research improves the effectiveness of the research [32, 33]. In areas of healthcare such as chronic illness treatment, having patients be the managers of the long-term goals of their care has shown significant improvements in patient wellness [34]. This also provides the opportunity to center the voices of marginalized groups who frequently experience discrimination within the healthcare process [35]. This is especially important in the development of healthcare AI where harmful biases can have significant impact on marginalized groups [36–38].

Patient and physician involvement in AI research happens at three stages: pre-, during, and post-development.

1. **Pre-development stage:** Pre-development is the conceptualization of a healthcare study, and is a challenging part of ethically and efficiently implementing patient involvement. However, pre-development is where methods and lessons from non-AI research with patient involvement can be most directly translated into a healthcare AI study. Patients can be involved either as a participant in a study, or more heavily involved as a contributor where they can voice their needs which can then inform development and use of health care AI. Patient important outcomes determined in consultation with patients can be incorporated as metrics to be optimized. Incorporating patient involvement must also incorporate ethical considerations that are informed by institutional ethics board principles, such as accountability, beneficence, confidentiality, dignity, transparency, rights etc. [38]. These ethical considerations must also be constantly challenged and reconsidered at every stage of the study.

    There is no doubt that patient involvement at the conceptual stage is very challenging. It entails a significant labour investment on the part of researchers and poor study design will mean wasted labour and a lack of respect for the involvement of patients [40]. When and how to make this type of investment is a complex question, as the benefits of patient involvement at the pre-development stage needs to be balanced against research efficiency and sustainability, in the context of limited public funding for research.

2. **Development stage:** Providers can be involved by aiding researchers with data annotation or by reviewing results. In terms of patient involvement, Banerjee et al. [41] propose the creation of a research advisory group made up of individuals who live with or care for someone with, a disorder that is observed for a given study. This group is then consulted throughout the development of the study for advice and feedback. This promotes patient trust, as they are able and encouraged to question the research at hand and share their concerns with the responsible party.
    Patients and providers are currently involved in AI research through the sharing of their data. However, healthcare providers have control over the data of their patients, and while providers are aware of the type of studies the data is sent to, patients typically are not. To prioritize patient involvement at this stage, information should be shared with the patients about their participation in a given study. For example, researchers requesting data from Canadian Primary Care Sentinel Surveillance Network (CPCSSN+), a Canadian network of primary care electronic medical records, must send an opt-out letter to physicians explaining the purposes of their study and proposed outcomes [42]. This letter could additionally be shared with patients by the physician, while maintaining patient confidentiality and ensuring that patients are informed, though this added step would

require substantial additional resources. If they then decide that they do not wish to participate in the study, per Personal Health Information Protection Act (PHIPA) law, patients would have the ability to opt out [43]. Protection of a patient's privacy also applies if they are deceased, where their legally authorized representative can choose to withdraw their information.

There are lots of gaps in current frameworks around ethical and privacy challenges to patients consenting to data sharing for secondary uses such as training AI [44]. Often patients are put in a place where they must consent to secondary uses of their data to receive primary care. However, if patients believe in the reciprocity, lack of exploitation and commitment to public good by the researchers, many choose to offer their data in order to potentially help future patients [45]. Effective patient involvement must allow patients the information and autonomy to make decisions on their data usage. At the same time consent frameworks must not be so restrictive as to under-include vulnerable population groups [45].

3. **Post-development stage:** Patient and provider involvement can be encouraged by testing AI applications in real clinical settings. This allows all parties involved to see how a model performs and grants the ability of patients and physicians to provide feedback to the research group, giving them the opportunity to improve their product.

Medical AI will only achieve its primary goal of improving healthcare measures are in place to ensure clinical relevance. This means including metrics that are relevant in clinical settings, evaluating clinical performance in clinical trials, and conducting prospective studies. Until then, these applications will not have much chance of crossing the AI Chasm or offering true benefit to their primary target: the patient.

## 2.2 Industry benefit

An analysis of arXiv, an online scientific paper archive, performed by Klinger et al. revealed that recent AI research is mostly driven by industries and organizations belonging to the private sector [46]. Similarly, Abdalla and Abdalla found that the majority of academic researchers in general AI research have had funding from large corporations [47]. When studies are industry-funded, conflicts of interest can frequently occur. Corporations have a fiduciary responsibility to their shareholders and therefore are expected to prioritize profit and reputation over objectives like social good or patient benefit. Researchers are motivated to seek industry funding because state-of-the-art AI research is costly and post-secondary education is being systematically de-funded [48]. The pool of grants are smaller and more competitive to apply for, and it is more and more attractive for researchers to accept private industry funding. For partnerships between industry and medical research to produce translatable products, a balance between patient and industry benefits must be established, while considering relevant ethical principles. A study that only benefits one of the parties would fail to translate. Table 2 summarizes criteria industry and researchers can consider to ensure the balance is addressed.

Because patients are in a vulnerable state, accountability is required by physicians. With AI, accountability is challenging when the AI model operates as a "black box". This term refers to how it can often be difficult to know how an AI reaches conclusions from given inputs since the decision process is hidden in parameters, uninterpretable by humans. Since the intent of medical AI is for it to be used alongside humans (i.e. not to replace medical professionals), knowing how a system arrives at a conclusion is crucial for the physician to explain diagnoses to the patient as well as to answer any questions that may arise from the given conclusion [49]. Similarly, if a physician presented their patient with a diagnosis without any explanation of their thought process and analysis, the patient might be inclined not to trust the diagnosis. Explanations are often not available from black box AI systems. To provide a comparable sense of reliability and trustworthiness that human physicians achieve at the bedside, a similar approach must be taken with AI, through explainable algorithmic decisions [49]. Systems that are credible and trusted by both patients and clinicians will also be more easily translated into care settings [14].

*Table 2* AI development requirements that ensure benefit to both patients and industry

| Requirements | Relevant ethical principles | Description |
|---|---|---|
| Developing explainable systems | *Trust:* Clinicians and patients are more likely to trust a system that they can understand and whose decisions they can verify. Additionally, in the clinic, physicians will use a system that they trust over one that they do not. | Systems and decisions that are not explained clearly are less likely to be trusted. Explainable systems allow healthcare practitioners to interpret results better, increasing reliability. |
| Prioritizing reputation among industry partners | *Beneficence:* Having a good reputation proves to the public that a company has the right intentions and wants to provide benefits.<br>*Non-maleficence:* Having a good reputation shows the public that a company is avoiding harm.<br>*Trust*: The public is more likely to trust a company with a good reputation. | A bad reputation can hamper the success of products. A good reputation may lead to higher sales which leads to more product development, benefiting both the corporation and the patients. |
| Overall healthcare benefits | *Beneficence:* Creating applications that offer overall healthcare benefits still provides benefits to individual patients since they are part of the system.<br>*Justice:* Focusing on the healthcare system as a whole promotes the concept of fair and equal care. | Applications that benefit healthcare as a whole (e.g. smart scheduling) are a way for corporations to avoid the legal and ethical implications of patient-involved technologies. |

Releasing effective and trustworthy patient-centred applications also provides potential advantages to industry partners involved in building healthcare applications. From a commercial point of view, if a product performs well and is trusted by stakeholders, products of the same type are more likely to be released, leading to increased sales and/or contract renewals, ultimately strengthening the product development cycle. A similar thought process can be applied to AI algorithms. If patients and clinicians trust algorithms, a company's reputation will be improved, leading to more business opportunities and potentially lucrative collaborations [15]. This can also go the other way where untrustworthy applications and their creators will lose business, funding, and reputation.

Overall healthcare benefit is also in the interest of industry. In addition, improving workflow and efficiency in clinical settings are also valid objectives of AI. Improvements to overall care pathways, for example by offering smart scheduling for physicians and residents, also contribute to overall healthcare benefit [51]. Another example is creating detailed repositories of medical language information [52]. These types of applications are a way for industries to partially escape the patient side of the trade-off. They can create useful applications that do not carry the underlying legal and ethical implications that usually follow patient-involved technologies. AI adoption in healthcare also risks increasing inequality in access to healthcare if access to these tools is not available to all patients and healthcare practitioners. Healthcare has already demonstrated a lack of distributive justice [51] and AI can exacerbate this issue if not handled with care. Nonetheless, applications of these sorts have downstream benefits for patients in the form of lower costs and quicker triage which ultimately improves the healthcare system.

There are frameworks present for industries to use as basis for equitable health AI, such as the framework for equitable innovation in health and medicine by US National Academy of Sciences [52] and blueprint for trustworthy AI by Coalition for Health AI [53]. These function as "soft law" providing companies with structure for equitable and responsible health innovation without them being legally bound to adhere to them. Further developing these frameworks in the context of improving trustworthiness, improving overall healthcare, and thus providing benefit to the industry is crucial in having them be a part of regular industrial practice.

## 2.3 Profit and funding

Funding is of critical importance in all kinds of AI research, as the necessary technology and equipment can be costly. Two main sources of funding can be identified: public (government) and private (industry). Public sources of funding have the mandate of increasing the welfare of the general population and are in place to promote research with clear impacts for the public [56].

However, the limited funds in government supported research means researchers have to face stiff competition for grants. Many are forced to turn to private sector funding as a result. Industry funders are often willing to endorse riskier proposals, with the idea that high-risk investments can bring high rewards [56]. They also provide larger sums of money. According to Statistics Canada, in 2024 the total funding for natural sciences and engineering research was $54 billion [57]. Table 3 summarizes the distribution of these funds by funder and recipient. There is a large difference in total amounts provided, and while the majority of public funding is distributed to outside parties such as academic institutions and grant recipients, private funding mostly stays within that sector. It should be noted that internal industry funding is accessed differently than public sector grants, and the lack of availability of research funding provided by the private sector to researchers further exacerbates the divide between public sector and private sector research funding.

| Recipients of Research Funding | Funder: Canadian Provincial and Federal Governments (in billion CAD) | Funder: Private Industry (in billion CAD) | Funder: Higher education sector (in billion CAD) |
|---|---|---|---|
| Federal and provincial government institutions | 3.19 | | |
| Higher education and academic sector | 5.79 | 1.2 | 9.1 |
| Private industry | 2.16 | 25.1 | |
| Total Funding | 11.32 | 26.4 | 9.1 |

*Table 3: Recipients of research funding in Canada in 2024 in billion Canadian dollars according to Statistics Canada [49]*

From the limited data available regarding public and private healthcare funding, similar trends can also be observed. In the US private funding in 2005 accounted for two-thirds of total healthcare research funding, with public funding accounting for one-third [58]. Public funding in US healthcare has been more focused on knowledge creation and assessing long term impacts on patients, whereas private funding is focused more on commercial applications [58].

The large amount of money behind privately funded research is concerning because industries are looking for net profits on the produced work. In medical research, that means generating profit from patients, which may conflict with the goal of patient benefit. Industries can be incentivized to mislead patients about the effects of a treatment, such as the anti arthritic drug Vioxx where the manufacturers used misleading research practices to supress knowledge of the negative effects of the drug [59]. Academic medicine has also been the subject of controversy for physicians receiving financial incentives to recruit patients into clinical trials, creating a conflict of interest in carrying out clinical trials [60–62]. AI applications and algorithms could likewise be influenced to recommend unnecessary tests or expensive medication, ultimately harming patients and healthcare [62].

There are two related concepts to consider: transparency at the level of the machine's decisions and at the level of industry processes. A partial solution to the problem of transparency would be to produce open-source algorithms, which can be reviewed and verified by outside parties to ensure no harm is caused by hidden decisions [63]. This addresses transparency in the algorithms themselves. Open-source software may also be partially corrective to the lack of transparency in industry motivations and how they are balancing profits with patient benefits. However, reviewing source code is complex, which highlights the need for pragmatic randomized clinical trials. Before a new tool can be integrated into care, there must first be clear evidence that the technology provides a net health benefit that outweighs the harms associated with profit.

Creating open technologies promotes a trust-based relationship between industry and patients as it allows patients to (at least in principle) check that they are not being harmed. Additionally, constructing beneficial and transparent algorithms instead of solely aiming to produce revenue would not only reduce the scientific waste mentioned in [11], but in the long run result in additional profit. If results are proven to be clinically relevant and useful, the algorithm is more likely to be used in real clinical situations which would ultimately generate the profit that industries are looking for.

## 3 Conclusion

The future integration of AI into clinical settings seems inevitable. While it has the potential to benefit all parties involved, the patient and their rights could be dismissed while industry prioritizes revenue. The patient-industry trade-off is that industry benefits will not always align with patient benefits. We have seen this before in medicine, where pharmaceutical companies are known to invest in advertising to convince healthy people they need treatments [63], leading to benefits for companies over patients. A version of this conflict of interest can already be seen in AI technologies in healthcare.

Patient benefits can be centered in AI development using clinically relevant metrics and involving metrics that patients find the most valuable. Validation with evidence allows patients and clinicians to have more trust when using AI models in healthcare. Co-creation with patients and physicians directly involved in the development of AI models would also allow AI models to directly center the needs of patients and physicians. Working towards these benefits are at direct odds with industry strategies of maximizing short-term profitability which is a significant conflict of interest with promoting patient benefit.

To address this trade-off, this paper introduces the following guidelines: modelling AI studies after medicine by including clinically relevant metrics, validating results with real world evidence, promoting patient and physician involvement, building trustworthy and explainable systems, fostering positive corporate reputation, providing overall healthcare benefits, and being transparent about source of funding. These methods are essential not only to ensure a balance between patient and industry but are also necessary for the implementation of algorithms into real-world clinical settings. The sooner these are addressed in AI literature and practice, the sooner we will see these types of technologies working alongside our physicians, where they can have real world impact.

In context of existing research to provide guidance towards responsible and ethical medical AI development [15, 52, 53], the short-term goal should be to further develop these guidelines to adequately address the trade-off with respect to patient benefit. These practices can be more concretely assessed through implementation in real world AI development. This will require healthcare practitioners and AI developers to commit to studies that assess real world evidence on patient benefits. The long-term goal of this is to provide regulatory frameworks on medical AI development that prioritize patient benefit, like how existing frameworks such as Institutional Review Boards (IRB), clinical regulations and privacy laws are intended to protect patients.

## Compliance with Ethical Standards

We, the authors, have received funding from external bodies that may prove to be a potential conflict of interest. This research has received funding from National Institutes of Health, USA under a grant for advancing the ethical development of AI. The research has also received funding from Connected Minds program, Canada, supported by Canada First Research Excellence Fund. No human or animal participants were involved in this study. This manuscript represents original research work by the authors that has not been published or submitted elsewhere.

## References

[1] Mitsala, A., Tsalikidis, C., Pitiakoudis, M., Simopoulos, C., Tsaroucha, A.K.: Artificial Intelligence in Colorectal Cancer Screening, Diagnosis and Treatment. A New Era. Current Oncology **28**(3), 1581–1607 (2021) https://doi.org/10.3390/ curroncol28030149

[2] Deshmukh, P.B., Kashyap, K.L.: Solution Approaches for Breast Cancer Classification Through Medical Imaging Modalities Using Artificial Intelligence. In: Zhang, Y.-D., Senjyu, T., So-In, C., Joshi, A. (eds.) Smart Trends in Computing and Communications. Lecture Notes in Networks and Systems, pp. 639–651. Springer, Singapore (2022). https://doi.org/10.1007/978-981-16-4016-2 61

[3] Bhandary, A., G, A.P., Basthikodi, M., M, C.K.: Early Diagnosis of Lung Cancer Using Computer Aided Detection via Lung Segmentation Approach. International Journal of Engineering Trends and Technology **69**(5), 85–93 (2021) https://doi.org/10.14445/22315381/IJETT-V69I5P213 arXiv:2107.12205

[4] Chanu, M.M., Thongam, K.: Computer-aided detection of brain tumor from magnetic resonance images using deep learning network. Journal of Ambient Intelligence and

Humanized Computing **12**(7), 6911–6922 (2021) https://doi.org/10. 1007/s12652-020-02336-w

[5] Saygılı, A.: A new approach for computer-aided detection of coronavirus (COVID19) from CT and X-ray images using machine learning methods. Applied Soft Computing **105**, 107323 (2021) https://doi.org/10.1016/j.asoc.2021.107323

[6] Iacoviello, M., Santamato, V., Pagano, A., Marengo, A.: "Interpretable AI-driven multi-objective risk prediction in heart failure patients with thyroid dysfunction," Frontiers in Digital Health, 12;7 :1583399. (2025), https://doi.org/10.3389/fdgth.2025.1583399

[7] Uzuner, O., Solti, I., Cadag, E.: Extracting medication information from clinical¨ text. Journal of the American Medical Informatics Association **17**(5), 514–518 (2010) https://doi.org/10.1136/jamia.2010.003947

[8] Tom, E., Keane, P.A., Blazes, M., Pasquale, L.R., Chiang, M.F., Lee, A.Y., Lee, C.S., and AAO Artificial Intelligence Task Force: Protecting Data Privacy in the Age of AI-Enabled Ophthalmology. Translational Vision Science & Technology **9**(2), 36 (2020) https://doi.org/10.1167/tvst.9.2.36

[9] Vellido, A.: Societal Issues Concerning the Application of Artificial Intelligence in Medicine. Kidney Diseases **5**(1), 11–17 (2019) https://doi.org/10.1159/ 000492428

[10] Vrudhula, A., Kwan, A.C., Ouyang, D., Cheng, S.: Machine Learning and Bias in Medical Imaging: Opportunities and Challenges. Circulation: Cardiovascular Imaging **17**(2), 015495 (2024) https://doi.org/10.1161/CIRCIMAGING.123. 015495 . Publisher: American Heart Association. Accessed 2025-02-07

[11] Chalmers, I., Glasziou, P.: Avoidable waste in the production and reporting of research evidence. The Lancet **374**(9683), 86–89 (2009) https://doi.org/10.1016/ S0140-6736(09)60329-9 . Publisher: Elsevier. Accessed 2022-03-09

[12] The Lancet.: What is the purpose of medical research? The Lancet **381**(9864), 347 (2013) https://doi.org/10.1016/S0140-6736(13)60149-X . Publisher: Elsevier. Accessed 2022-03-09

[13] Keane, P.A., Topol, E.J.: With an eye to AI and autonomous diagnosis. npj Digital Medicine **1**(1), 1–3 (2018) https://doi.org/10.1038/s41746-018-0048-y . Number: 1 Publisher: Nature Publishing Group. Accessed 2022-03-11

[14] Kelly, C.J., Karthikesalingam, A., Suleyman, M., Corrado, G., King, D.: Key challenges for delivering clinical impact with artificial intelligence. BMC Medicine **17**(1), 195 (2019) https://doi.org/10.1186/s12916-019-1426-2 . Accessed 2022-0311 [1]

[15] Adler-Milstein, J., Aggarwal, N., Ahmed, M., Castner, J., Evans, B.J., Gonzalez, A.A., James, C.A., Lin, S., Mandl, K.D., Matheny, M.E., Sendak, M.P.: "Meeting the Moment: Addressing Barriers and Facilitating Clinical Adoption of Artificial Intelligence in Medical Diagnosis - NAM," National Academy of Medicine. (2022). Accessed: Oct. 07, 2025. [Online]. Available: https://nam.edu/perspectives/meeting-the-moment-addressing-barriers-and-facilitating-clinical-adoption-of-artificial-intelligence-in-medical-diagnosis/

[16] The National Commission for the Protection of Human Subjects of Biomedical and Behavioral Research: The Belmont Report (1979). https://www.hhs.gov/ohrp/regulations-and-policy/belmont-report/read-the-belmont-report/index.html Accessed 2024-04-25

[17] Guyatt, G., Montori, V., Devereaux, P.J., Schünemann, H., Bhandari, M.: Patients at the center: In our practice, and in our use of language. ACP Journal Club **140**(1), 11 (2004) https://doi.org/10.7326/ACPJC-2004-140-1-A11 . Publisher: American College of Physicians. Accessed 2025-04-09

[18] Gandhi, G.Y., Murad, M.H., Fujiyoshi, A., Mullan, R.J., Flynn, D.N., Elamin, M.B., Swiglo, B.A., Isley, W.L., Guyatt, G.H., Montori, V.M.: Patient-Important Outcomes in Registered Diabetes Trials. JAMA **299**(21), 2543–2549 (2008) https://doi.org/10.1001/jama.299.21.2543 . Accessed 2025-04-10

[19] Handelman, G.S., Kok, H.K., Chandra, R.V., Razavi, A.H., Huang, S., Brooks, M., Lee, M.J., Asadi, H.: Peering Into the Black Box of Artificial Intelligence: Evaluation Metrics of Machine Learning Methods. American Journal of Roentgenology **212**(1), 38–43 (2019) https://doi.org/10.2214/AJR.18.20224 . Publisher: American Roentgen Ray Society. Accessed 2022-03-11

[20] Shah, N.H., Milstein, A., Bagley, P. Steven C.: Making Machine Learning Models Clinically Useful. JAMA **322**(14), 1351–1352 (2019) https://doi.org/10.1001/jama.2019.10306 . Accessed 2022-03-11

[21] Halligan, S., Altman, D.G., Mallett, S.: Disadvantages of using the area under the receiver operating characteristic curve to assess imaging tests: A discussion and proposal for an alternative approach. European Radiology **25**(4), 932–939 (2015) https://doi.org/10.1007/s00330-014-3487-0

[22] Guthrie, S., Krapels, J., Lichten, C., Wooding, S.: 100 Metrics to Assess and Communicate the Value of Biomedical Research: An Ideas Book. RAND Corporation (2016). https://doi.org/10.7249/RR1606

[23] Booth, C.M., Eisenhauer, E.A., Gyawali, B., Tannock, I.F.: Progression-Free Survival Should Not Be Used as a Primary End Point for Registration of Anticancer Drugs.

Journal of Clinical Oncology **41**(32), 4968–4972 (2023) https://doi.org/10.1200/JCO.23.01423 . Publisher: Wolters Kluwer. Accessed 2024-04-25

[24] Angus, D.C.: Randomized Clinical Trials of Artificial Intelligence. JAMA **323**(11), 1043–1045 (2020) https://doi.org/10.1001/jama.2020.1039 . Accessed 2022-03-20

[25] Miller, R.A.: Computer-assisted diagnostic decision support: History, challenges, and possible paths forward. Advances in Health Sciences Education **14**(1), 89–106 (2009) https://doi.org/10.1007/s10459-009-9186-y

[26] Tang, A., Tam, R., Cadrin-Chênevert, A., Guest, W., Chong, J., Barfett, J., Chepelev, L., Cairns, R., Mitchell, J.R., Cicero, M.D., Poudrette, M.G., Jaremko, J.L., Reinhold, C., Gallix, B., Gray, B., Geis, R., O'Connell, T., Babyn, P., Koff, D., Ferguson, D., Derkatch, S., Bilbily, A., Shabana, W.: Canadian Association of Radiologists White Paper on Artificial Intelligence in Radiology. Canadian Association of Radiologists Journal **69**(2), 120–135 (2018) https://doi.org/10. 1016/j.carj.2018.02.002

[27] US Center for Food and Drug Administration: Computer-Assisted Detection Devices Applied to Radiology Images and Radiology Device Data - Premarket Notification [510(k)] Submissions. FDA (Tue, 09/27/2022 - 17:13)

[28] Zhang, Y., Shi, J., Peng, Y., Zhao, Z., Zheng, Q., Wang, Z., et al.: Artificial intelligence-enabled screening for diabetic retinopathy: a real-world, multicenter and prospective study, BMJ Open Diabetes Research & Care 8(1). (2020). https://doi.org/10.1136/bmjdrc-2020-001596.

[29] Stinson, C.: Healthy Data: Policy solutions for big data and AI innovation in health (2018). https://utoronto.scholaris.ca/server/api/core/bitstreams/ 3e4bc50f-084a-4fc3-8eb1-59ea5dd66209/content

[30] Litva, A., Coast, J., Donovan, J., Eyles, J., Shepherd, M., Tacchi, J., Abelson, J., Morgan, K.: 'The public is too subjective': public involvement at different levels of health-care decision making. Social Science & Medicine **54**(12), 1825–1837 (2002) https://doi.org/10.1016/S0277-9536(01)00151-4 . Accessed 2025-04-24

[31] Saini, P., Hassan, S.M., Morasae, E.K., Goodall, M., Giebel, C., Ahmed, S., Pearson, A., Harper, L.M., Cloke, J., Irvine, J., Gabbay, M.: The value of involving patients and public in health services research and evaluation: A qualitative study. Research Involvement and Engagement **7**(1), 49 (2021) https://doi.org/10.1186/ s40900-021-00289-8

[32] Forsythe, L., Heckert, A., Margolis, M.K., Schrandt, S., Frank, L.: Methods and impact of engagement in research, from theory to practice and back again: early findings from the Patient-Centered Outcomes Research Institute. Quality of Life Research **27**(1), 17–31 (2018) https://doi.org/10.1007/s11136-017-1581-x . Accessed 2025-05-07

[33] Boivin, A., Richards, T., Forsythe, L., Grégoire, A., L'Espérance, A., Abelson, J., Carman, K.L.: Evaluating patient and public involvement in research. BMJ **363**, 5147 (2018) https://doi.org/10.1136/bmj.k5147 . Publisher: British Medical Journal Publishing Group Section: Editorial. Accessed 2025-05-07

[34] Kadu, M.K., Stolee, P.: Facilitators and barriers of implementing the chronic care model in primary care: a systematic review. BMC Family Practice **16**(1), 12 (2015) https://doi.org/10.1186/s12875-014-0219-0 . Accessed 2025-05-07

[35] Rosen, L.T.: Mapping out epistemic justice in the clinical space: using narrative techniques to affirm patients as knowers. Philosophy, Ethics, and Humanities in Medicine **16**(1), 9 (2021) https://doi.org/10.1186/s13010-021-00110-0 . Accessed 2025-05-07

[36] Seyyed-Kalantari, L., Zhang, H., McDermott, M.B.A., Chen, I.Y., Ghassemi, M.: Underdiagnosis bias of artificial intelligence algorithms applied to chest radiographs in under-served patient populations. Nature Medicine **27**(12), 2176–2182 (2021) https://doi.org/10.1038/s41591-021-01595-0 . Publisher: Nature Publishing Group. Accessed 2024-09-30

[37] Estiri, H., Strasser, Z.H., Rashidian, S., Klann, J.G., Wagholikar, K.B., McCoy, T.H. Jr, Murphy, S.N.: An objective framework for evaluating unrecognized bias in medical AI models predicting COVID-19 outcomes. Journal of the American Medical Informatics Association **29**(8), 1334–1341 (2022) https://doi.org/10.1093/jamia/ocac070 . Accessed 2024-07-03

[38] Banerjee, I., Bhimireddy, A.R., Burns, J.L., Celi, L.A., Chen, L.-C., Correa, R., Dullerud, N., Ghassemi, M., Huang, S.-C., Kuo, P.-C., Lungren, M.P., Palmer, L., Price, B.J., Purkayastha, S., Pyrros, A., Oakden-Rayner, L., Okechukwu, C., Seyyed-Kalantari, L., Trivedi, H., Wang, R., Zaiman, Z., Zhang, H., Gichoya, J.W.: Reading Race: AI Recognises Patient's Racial Identity In Medical Images. The Lancet Digital Health **4**(6), 406–414 (2022) https://doi.org/10.1016/S2589-7500(22)00063-2 . arXiv:2107.10356 [cs, eess]. Accessed 2023-10-27

[39] Minogue, V., Salsberg, J.: Meaningful and Safe: The Ethics and Ethical Implications of Patient and Public Involvement in Health And Medical Research. Ethics International Press Limited, Bradford, UNITED KINGDOM (2024). Chap. 2. http://ebookcentral.proquest.com/lib/queen-ebooks/detail.action?docID=31731259 Accessed 2025-04-27

[40] Minogue, V., Salsberg, J.: Meaningful and Safe: The Ethics and Ethical Implications of Patient and Public Involvement in Health And Medical Research, pp. 3–4. Ethics International Press Limited, Bradford, UNITED KINGDOM (2024). Chap. 1. http://ebookcentral.proquest.com/lib/queen-ebooks/detail.action?docID=31731259 Accessed 2025-04-27

[41] Banerjee, S., Alsop, P., Jones, L., Cardinal, R.N.: Patient and public involvement to build trust in artificial intelligence: A framework, tools, and case studies. Patterns **3**(6), 100506 (2022) https://doi.org/10.1016/j.patter.2022.100506

[42] Network, C.P.C.S.S.: Canadian Primary Care Sentinel Surveillance Network (CPCSSN) - About Us. Accessed November 10, 2022

[43] Personal Health Information Protection Act, S.O. 2004, c. 3, Sched. A. https://www.ontario.ca/laws/statute/04p03#top (2004).

[44] Moulaei K, Akhlaghpour S, Fatehi F.: Patient consent for the secondary use of health data in artificial intelligence (AI) models: A scoping review. International Journal of Medical Informatics. 1(198), 105872 (2025).

[45] Evans BJ, Bihorac A.: Co-creating Consent for Data Use — AI-Powered Ethics for Biomedical AI. NEJM AI. 1(7), (2024).

[46] Klinger, J., Mateos-Garcia, J.C., Stathoulopoulos, K.: A Narrowing of AI Research? SSRN Scholarly Paper ID 3698698, Social Science Research Network, Rochester, NY (September 2020). https://doi.org/10.2139/ssrn.3698698

[47] Abdalla, M., Abdalla, M.: The Grey Hoodie Project: Big Tobacco, Big Tech, and the threat on academic integrity. In: Proceedings of the 2021 AAAI/ACM Conference on AI, Ethics, and Society, pp. 287–297 (2021). https://doi.org/10.1145/3461702.3462563

[48] Romard, R.: Whoever wins the election, Ontario must reinvest in post-secondary education - CCPA (2025). https://www.policyalternatives.ca/news-research/whoever-wins-the-election-ontario-must-reinvest-in-post-secondary-education/ Accessed 2025-06-20

[49] Goisauf, M., Cano Abad´ıa, M.: Ethics of AI in Radiology: A Review of Ethical and Societal Implications. Frontiers in Big Data **5** (2022)

[50] Tjoa, E., Guan, C.: A Survey on Explainable Artificial Intelligence (XAI): Toward Medical XAI. IEEE Transactions on Neural Networks and Learning Systems **32**(11), 4793–4813 (2021) https://doi.org/10.1109/TNNLS.2020.3027314 . Conference Name: IEEE Transactions on Neural Networks and Learning Systems1.

[51] Daniels N.: Justice, Health, and Healthcare. The American Journal of Bioethics. 1(2):2–16. (2001)

[52] National Academies of Sciences, Engineering, and Medicine. Toward equitable innovation in health and medicine: a framework. (2023)

[53] Coalition for Health AI T. Blueprint for trustworthy AI implementation guidance and assurance for healthcare. (2023)

[54] Perelstein, E., Rose, A., Hong, Y.-C., Cohn, A., Long, M.T.: Automation Improves Schedule Quality and Increases Scheduling Efficiency for Residents. Journal of Graduate Medical Education **8**(1), 45–49 (2016) https://doi.org/10. 4300/JGME-D-15-00154.1 . Accessed 2022-03-21

[55] Bodenreider, O.: The Unified Medical Language System (UMLS): integrating biomedical terminology. Nucleic Acids Research **32**(suppl 1), 267–270 (2004) https://doi.org/10.1093/nar/gkh061 . Accessed 2022-03-21

[56] Government Grant Fund Vs. Private Grant Fund – Which is Better? – Conduct Science. https://conductscience.com/ government-grant-fund-vs-private-grant-fund-which-is-better/ Accessed 2022-03-14

[57] Canada, S.C.: Gross domestic expenditures on research and development, by science type and by funder and performer sector (2024). https://www150.statcan.gc.ca/t1/tbl1/en/tv.action?pid=2710027301 Accessed 2025-05-06

[58] Sampat, B.N.: THE IMPACT OF PUBLICLY FUNDED BIOMEDICAL AND HEALTH RESEARCH: A REVIEW. In: Measuring the Impacts of Federal Investments in Research: A Workshop Summary. National Academies Press (US) (2011). https://www.ncbi.nlm.nih.gov/ books/NBK83123/ Accessed 2025-05-06

[59] Faunce, T.A.: The vioxx pharmaceutical scandal: Peterson v. merke sharpe & dohme (aust) pty ltd (2010) 184 fcr 1. Journal of law and medicine **18**, 38–49 (2010)

[60] Chabner, B.A., Bates, S.E.: Conflict of Interest: An Ethical Firestorm with Consequences for Cancer Research. The Oncologist **23**(12), 1391–1393 (2018) https://doi.org/10.1634/theoncologist.2018-0662 . Accessed 2024-04-26

[61] Prasad, V., Rajkumar, S.V.: Conflict of interest in academic oncology: moving beyond the blame game and forging a path forward. Blood Cancer Journal **6**(11), 489 (2016) https://doi.org/10.1038/bcj.2016.101 . Accessed 2024-04-26

[62] Wayant, C., Turner, E., Meyer, C., Sinnett, P., Vassar, M.: Financial Conflicts of Interest Among Oncologist Authors of Reports of Clinical Drug Trials. JAMA Oncology **4**(10), 1426–1428 (2018) https://doi.org/10.1001/jamaoncol.2018.3738 . Accessed 2024-04-26

[63] Moynihan, R., Heath, I., Henry, D.: Selling sickness: the pharmaceutical industry and disease mongering. BMJ : British Medical Journal **324**(7342), 886–891 (2002). Accessed 2024-04-30